\documentclass[%
 reprint,
 amsmath,amssymb,
 aps,
]{revtex4-2}

\usepackage{graphicx}% Include figure files
\usepackage{dcolumn}% Align table columns on decimal point
\usepackage{bm}% bold math
\usepackage{xcolor}
\begin{document}

\preprint{APS/123-QED}

\title{Curvature Effect on the Speed of Sound}% Force line breaks with \\
%\thanks{Speed of Sound}%

\author{Anshuman Verma}
 \email{anshuman18@iiserb.ac.in}
% \altaffiliation[at ]{IISER Bhopal}%Lines break automatically or can be forced with \\
\author{Asim Kumar Saha}%
 \email{asim21@iiserb.ac.in}
\author{Ritam Mallick}%
 \email{mallick.at.iiserb.ac.in}
\affiliation{%
Indian Institute of Science Education and Research Bhopal\\
 India, 462066 
}%

% \collaboration{MUSO Collaboration}%\noaffiliation

%  \homepage{http://www.Second.institution.edu/~Charlie.Author}
% \affiliation{
%  Second institution and/or address\\
%  This line break forced% with \\
% }%
% \affiliation{
%  Third institution, the second for Charlie Author
% \affiliation{%
%  Authors' institution and/or address\\
%  This line break forced with \textbackslash\textbackslash
% }%

% \collaboration{CLEO Collaboration}%\noaffiliation

\date{\today}% It is always \today, today,

\begin{abstract}
The speed of sound refers to the rate at which information travels from one point to another. It is a positive quantity and bounded by causality. It is defined as the rate of change of pressure with respect to the system's density. In this article, we derive a covariant equation for the sound wave and demonstrate how the wave equation is modified in the general relativistic formalism. One can then define an effective speed of sound by attenuating the usual definition of sound speed with the gravitational metric potential. The general relativistic curvature effect is observed to reduce the speed of sound when computed inside a neutron star. This effectively makes the star relatively softer (according to the equation of state). The change in the effective sound speed can be easily visualised if one redefines the non-radial modes in terms of it. The modes do not change, but the space-time curvature reduces the amplitude of the oscillation modes. The formalism is suited for studying astrophysical compact objects.

\end{abstract}

\maketitle

%\tableofcontents

\section{Introduction}\label{introduction}

The propagation of a wave, be it transverse or longitudinal, is dependent on space-time (ST). It forms the basis of understanding the medium and the curvature of ST \citep{visser_2010,kees_2024, delory_2024}. In astrophysics, one primarily encounters two basic types of waves: electromagnetic waves and sound waves \citep{jones_2025}. Recently, gravitational waves have also been found to be a tool for probing astrophysical environments that were previously inaccessible \cite{Abbott_2017, Ozel_NS, Annala:2021gom, Malik_2024, Tang:2024jvs, Watts_NS, Haque_2024, Haque_2025, Prasad_2020, Singh_2021, Radice_2018, Most_2018, Most_2019, Most_2020, Weih_2020, Bauswein_2017}. The propagation of sound waves in astrophysics is primarily governed by hydrodynamics, which depends on both the properties of the medium and the ST curvature \cite{Tang:2024jvs, Fujimoto:2022ohj}. Therefore, it is expected that sound wave propagation in a dense astrophysical medium would be governed both by medium properties and the intrinsic ST curvature.

The longitudinal waves are mostly the energy or pressure perturbation (mechanical) of the medium, which results in either a sound wave or a shock wave \cite{liu_2011, prasad_2018,verma_2022, verma_2023, Tan:2021nat}.
Sound waves in astrophysics play an immense role, ranging from governing the formation of jets in astrophysical fluids to describing the properties of matter through the equation of state (EoS) \cite{bedaque_2015, Altiparmak:2022bke, Annala:2023cwx, sagnik_sos, Verma_2025, Tewari_2025, Ecker_2022}. It is also bounded by causality, expressing that information cannot travel faster than the speed of light. All these are of immense importance in astrophysics, particularly near a massive compact object where curvature properties are non-negligible. Sound wave propagation is then affected by ST curvatures \cite{avis_2020, Bi_2021, Handley_2019, sagnik_sos, Ecker_2022}. In this article, we aim to investigate the effects of curvature on sound wave propagation in and around a neutron star (NS).

The matter at extreme densities is still speculative, and one of the most interesting astrophysical laboratories is the NSs, where they exist in their innermost core \cite{Komoltsev:2021jzg, Komoltsev:2022bif, verma_2022,  Verma_2025}. At these densities, speculation ranges from superfluid neutrons and protons, pions and kaons, hyperons, to deconfined quarks \cite{Hebeler:2013nza, Hatfield_2021, verma_2023, Verma_2025}. Therefore, the EoS at these densities is an active area of research \cite{Lattimer_NS, Steiner_NS, Tan:2020ics}, and one of the constraining features of the EoS arises from the thermodynamic consistency of the speed of sound, which restricts it to be between 0 and 1 (assuming the speed of light to be 1) \cite{bedaque_2015,kanakis_2020, hippert_2021}. The relation between the EoS and the sound speed is given by $c_s^2=\frac{\partial p}{\partial \epsilon}$.

However, the commonly used definition is derived in a special relativistic framework \cite{moustakidis_2017, tews_2018}. Inside neutron stars, spacetime curvature is strong, and whether the same definition remains valid or requires general relativistic (GR) corrections has not been systematically investigated for neutron star matter. Existing studies have considered such corrections in cosmological \cite{pen_2016,hindmarsh_2018, hindmarsh_2019, hippert_2021, Wang_2022, tian_2025} or black hole contexts \cite{takahashi_1990,tenkanen_2022, Ning_2025}, but not within stellar interiors where curvature and pressure gradients are intertwined.

In this work, we derive the covariant formulation of pressure perturbations and obtain a general-relativistic expression for the speed of sound inside neutron stars. We systematically compare the special- and general-relativistic formulations, quantify curvature-induced modifications to $c_s$, and explore their implications using a range of realistic EoSs consistent with current observational constraints. Our results provide a quantitative framework for assessing the effects of spacetime curvature on the propagation of sound waves inside compact stars, thereby refining one of the fundamental quantities governing the dense-matter EoS \cite{Baym:1975mf, Altiparmak:2022bke}.

We begin by deriving the covariant form of the pressure (sound) wave equation in both special and general relativity. The paper is organised as follows: Section \ref{sec2} presents the derivation of the covariant sound-wave equation in special relativistic and general-relativistic backgrounds.  
Section \ref{sec6} presents and analyses the results, and Section \ref{Summary_and_Conclusion} concludes the study with a summary of key findings. Throughout this work, we adopt natural units with (G = c = $\hbar$ = 1).

\section{Formalism}\label{sec2}

The energy-momentum tensor ($T^{\mu \nu}$) for a perfect fluid has the form (without viscosity and heat conduction),

  \begin{align}\label{c1} 
       T^{\mu\nu} &= w u^{\mu} u^{\nu} + p g^{\mu \nu}  \nonumber \\
       \implies T^{\mu\nu} &= \epsilon u^{\mu} u^{\nu} + p k^{\mu \nu} 
  \end{align}
  where, enthalpy density ($w$) = pressure ($p$) + energy density ($\epsilon)$, $u^{\mu}$ are the fluid four velocity, $g^{\mu \mu}$ the metric tensor, $k^{\mu \nu}$ is the projection operator which projects quantities orthogonal to the fluid's motion(i.e, $k^{\mu \nu} u_{\mu} = 0$ with $u^{\mu} u_{\mu} = -1$), and is given by
  \begin{equation}
      k^{\mu \nu} = u^{\mu} u^{\nu} + g^{\mu \nu} 
  \end{equation}
Along with this, one also has the conservation of the energy-momentum tensor
\begin{equation}\label{c3}
    \nabla_{\mu}  T^{\mu\nu} = 0.
\end{equation}

The projection of the conservation of the energy-momentum (eq.\ref{c3}) along the world line of the fluid flow, i.e, along the four-velocity states 
\begin{align}\label{c4}
%  u_{\nu}\nabla_{\mu}  T^{\mu\nu} = 0 \nonumber \\
%  \implies u_{\nu} \nabla_{\mu}(\epsilon u^{\mu} u^{\nu} + p k^{\mu \nu}) = 0 \nonumber \\
%  \implies 
  u_{\nu} u^{\nu} \nabla_{\mu}(\epsilon u^{\mu} ) + p u_{\nu} \nabla_{\mu}(u^{\mu} u^{\nu} + g^{\mu \nu} ) = 0
\end{align}
The four-velocity and four-acceleration are always orthogonal, i.e their inner product vanishes 
 \begin{align}
    u_{\mu} a^{\mu} = u_{\mu} u^{\nu} \nabla_{\nu} u^{\mu} = 0 \nonumber 
\end{align}
This orthogonality is a direct consequence of the fact that the norm of the four-velocity is constant. Since the four-velocity's norm doesn't change, its derivative with respect to proper time must be zero, i.e,
 \begin{align}
 \frac{d}{d\tau}(u^{\mu} u_{\mu}) = u^{\nu} \nabla_{\nu}(u^{\mu} u_{\mu})= 0 \nonumber
\end{align}
implying that the four-velocity and four-acceleration are orthogonal. Using the given identity in Eqn. \ref{c4} with $\nabla_{\mu}g^{\mu \nu} = 0$, one obtains the conservation of energy-density
\begin{align}\label{c5}
 - \nabla_{\mu}(\epsilon u^{\mu} ) - p \nabla_{\mu} u^{\mu} = 0 \nonumber \\ 
%  \implies - \epsilon \nabla_{\mu} u^{\mu} - p \nabla_{\mu} u^{\mu} - u^{\mu} \nabla_{\mu}\epsilon = 0 \nonumber \\ 
%  \implies - (p + \epsilon) \nabla_{\mu} u^{\mu} - u^{\mu} \nabla_{\mu}\epsilon = 0 \nonumber \\
  \implies - (p + \epsilon) \nabla_{\mu} u^{\mu} = u^{\mu} \nabla_{\mu}\epsilon
\end{align}
%The above equation is known as the conservation of energy in general relativity.
Similarly, the momentum conservation equation is calculated using a projection operator,
 \begin{align}
     k^{\sigma}_{\nu}\nabla_{\mu} T^{\mu\nu} = 0 \nonumber \\
%     \implies k^{\sigma}_{\nu}\nabla_{\mu}(\epsilon u^{\mu} u^{\nu} + p k^{\mu \nu})=0 \nonumber \\
     \implies \epsilon k^{\sigma}_{\nu} u^{\mu} \nabla_{\mu} u^{\nu} + k^{\sigma \mu} \nabla_{\mu}p + p k^{\sigma}_{\nu} \nabla_{\mu}(u^{\mu}u^{\nu}) = 0 \nonumber
 \end{align}
and using the identity $ k^{\sigma}_{\nu} u^{\mu} \nabla_{\mu} u^{\nu} = u^{\mu} \nabla_{\mu} u^{\sigma}$ one obtains
\begin{align}\label{c6}
     (p+\epsilon)u^{\mu} \nabla_{\mu} u^{\sigma} = -k^{\sigma \mu} \nabla_{\mu}p
\end{align}

The covariant form of the wave equation can be derived from Eqn \ref{c5} and Eqn. \ref{c6} by taking the proper time derivative ($u^{\mu} \nabla_{\mu} = \frac{d}{d\tau}$) of energy conservation equation and covariant derivative ($\nabla_{\sigma}$) of momentum conservation equation respectively.

\begin{align}
    \frac{d^2\epsilon}{d\tau^2} = - (\epsilon + p)\nabla_{\mu} \frac{d}{d\tau} u^{\mu} \label{m1} \\
    (p+\epsilon)\nabla_{\sigma}\frac{d}{d\tau} u^{\sigma} = -k^{\sigma \mu} \nabla_{\sigma}\nabla_{\mu} p \label{m2}
\end{align}

Subtracting Eqn. \ref{m1} and Eqn. \ref{m2} one gets a wave equation
\begin{equation}
    \frac{d^2\epsilon}{d\tau^2} - k^{\sigma \mu} \nabla_{\sigma}\nabla_{\mu} p = 0 \label{m3}\\
\end{equation}

which could also be expressed as
\begin{equation}
   (u^{\mu} \nabla_{\mu})^2\epsilon - k^{\sigma \mu} \nabla_{\sigma}\nabla_{\mu} p = 0 \label{m4}. \\
\end{equation}

Having the perturbation to be of the form
\begin{align}
    p &= p + \delta p \nonumber \\
    \epsilon &= \epsilon + \delta \epsilon \nonumber \\
    u^{\mu} &= u^{\mu} + \delta u^{\mu} \nonumber
\end{align}

The wave equation takes the form
\begin{equation}\label{c7}
     \frac{1}{c_s^2}(u^{\mu} \nabla_{\mu})^2 \delta p - k^{\sigma \mu} \nabla_{\sigma}\nabla_{\mu} \delta p= 0
\end{equation}
where we are using the definition of the local speed of sound $c_s^2 = \left(\frac{\delta p }{\delta \epsilon}\right)_{s}$, and $u^{\mu}$. 

Simple transformation of the form 
\begin{align}
    g^{\mu\nu} \rightarrow \eta^{\mu\nu} \nonumber \\
    k^{\mu\nu} \rightarrow \eta^{\mu\nu} + u^{\mu} u^{\nu} \nonumber \\
    \nabla_{\mu} \rightarrow \partial_{\mu} \nonumber
\end{align}
where $\eta^{\mu \nu}$ is the flat-space metric, gives the special-relativistic form of the wave equation.

\begin{equation}\label{c8}
     \frac{1}{c_s^2}(u^{\mu} \partial_{\mu})^2 \delta p - k^{\sigma \mu} \partial_{\sigma}\partial_{\mu} \delta p= 0
\end{equation}

\subsection{Wave equation in spherically symmetric ST}

The wave equation can be studied in more detail if one assumes a static spherically symmetric metric given by

\begin{equation}
    ds^2= g_{\mu\nu}dx^{\mu}dx^{\nu}=- e^{2\psi(r)}dt^2 + e^{2\lambda(r)}dr^2 + r^{2}d\Omega^{2}
\end{equation}
where $d\Omega^{2} = d\theta^{2} + \sin^{2}({\theta}) d\phi^{2}$. 

The covariant derivative of $\delta p$ along $u^\mu$ is:
\[
u^\mu \nabla_\mu \delta p = e^{-\psi(r)} \frac{\partial \delta p}{\partial t}
\]

Thus, squaring this term and multiplying by $\frac{1}{c_s^2}$ one obtains

\begin{equation}\label{gr1}
    \frac{1}{c_s^2} (u^\mu \nabla_\mu)^2 \delta p = \frac{1}{c_s^2} e^{-2\psi(r)} \frac{\partial^2 \delta p}{\partial t^2}
\end{equation}

The covariant derivative $\nabla_\mu \nabla_\sigma \delta p$ is simply the partial derivative for a scalar field, thus simplifying the equations.

\begin{itemize}
  \item Radial derivative:
    \[
    k^{rr} \nabla_r \nabla_r \delta p = e^{-2\lambda(r)} \frac{\partial^2 \delta p}{\partial r^2} + \frac{e^{-2\lambda(r)}}{r^2} \partial_r \left( r^2 \partial_r \delta p \right)
    \]
  \item Angular derivatives:
    \[
    k^{\theta\theta} \nabla_\theta \nabla_\theta \delta p = \frac{1}{r^2} \left(\frac{\partial^2 \delta p}{\partial \theta^2} + \cot \theta \frac{\partial \delta p}{\partial \theta}\right)
    \]
    \[
    k^{\phi\phi} \nabla_\phi \nabla_\phi \delta p = \frac{1}{r^2 \sin^2 \theta} \frac{\partial^2 \delta p}{\partial \phi^2}
    \]
\end{itemize}

Since we are considering spherically symmetric pressure perturbations, $\delta p$ is independent of $\theta$ and $\phi$, so the angular derivatives vanish. Thus one has

\begin{equation}\label{gr2}
k^{\sigma\mu} \nabla_\sigma \nabla_\mu \delta p = e^{-2\lambda(r)} \left(\frac{\partial^2 \delta p}{\partial r^2} + \frac{2}{r} \frac{\partial \delta p}{\partial r}\right)
\end{equation}

The final wave equation now takes the form

\begin{equation}\label{grf}
\frac{1}{c_s^2} e^{-2\psi(r)} \frac{\partial^2 \delta p}{\partial t^2} - e^{-2\lambda(r)} \left(\frac{\partial^2 \delta p}{\partial r^2} + \frac{2}{r} \frac{\partial \delta p}{\partial r}\right) = 0
\end{equation}

This is the pressure wave or sound wave equation in the local rest frame of the fluid in a static spherically symmetric spacetime. The equation captures the propagation of sound waves in a relativistic fluid in a gravitational field.

For flat ST
\begin{align*}
    e^{-2\psi} \rightarrow 1\\
    e^{-2\lambda} \rightarrow 1\\
\end{align*}

The wave equation simplifies to the form

\begin{equation}\label{srf}
\frac{1}{c_s^2} \frac{\partial^2 \delta p}{\partial t^2} - \frac{\partial^2 \delta p}{\partial r^2} - \frac{2}{r} \frac{\partial \delta p}{\partial r} = 0
\end{equation}

\subsection{A particular solution}\label{sec4}

Consider a pressure perturbation in the radial direction having wavelength $\Lambda$ of the form

\begin{equation} \label{delp}
    \delta p(r,t) = \frac{h(t)}{r} \sin \bigg(\frac{2 \pi r}{\Lambda}\bigg) 
\end{equation}
%On using this form of perturbation in equation (\ref{grf}), we can find

Which reduced the wave equation in terms of $h(t)$
\begin{equation}
     \frac{\partial^2 h(t)}{\partial t^2} = -\Bigg(e^{(\psi(r) - \lambda(r)} c_s \frac{2 \pi}{\Lambda}\Bigg)^{2} h(t)
\end{equation}
implying $h(t)$ having a sinusoidal solution of effective frequency

\begin{equation*}
     \omega_{eff} = \Bigg(e^{(\psi(r) - \lambda(r)} c_s \frac{2 \pi}{\Lambda}\Bigg).
\end{equation*}

One can also define an effective sound speed as 
\begin{align*}
    ({c_{s}})_{eff} &= e^{(\psi(r) - \lambda(r))} c_s
\end{align*}

The exponential factor $e^{(\psi - \lambda)}$ encapsulates the influence of spacetime curvature (via $\lambda$) and gravitational redshift (via $\psi$), thus modifying the conventional sound speed definitions.

\section{Results}\label{sec6}

\begin{figure*}
    \centering
    \includegraphics[width=0.42\linewidth]
    {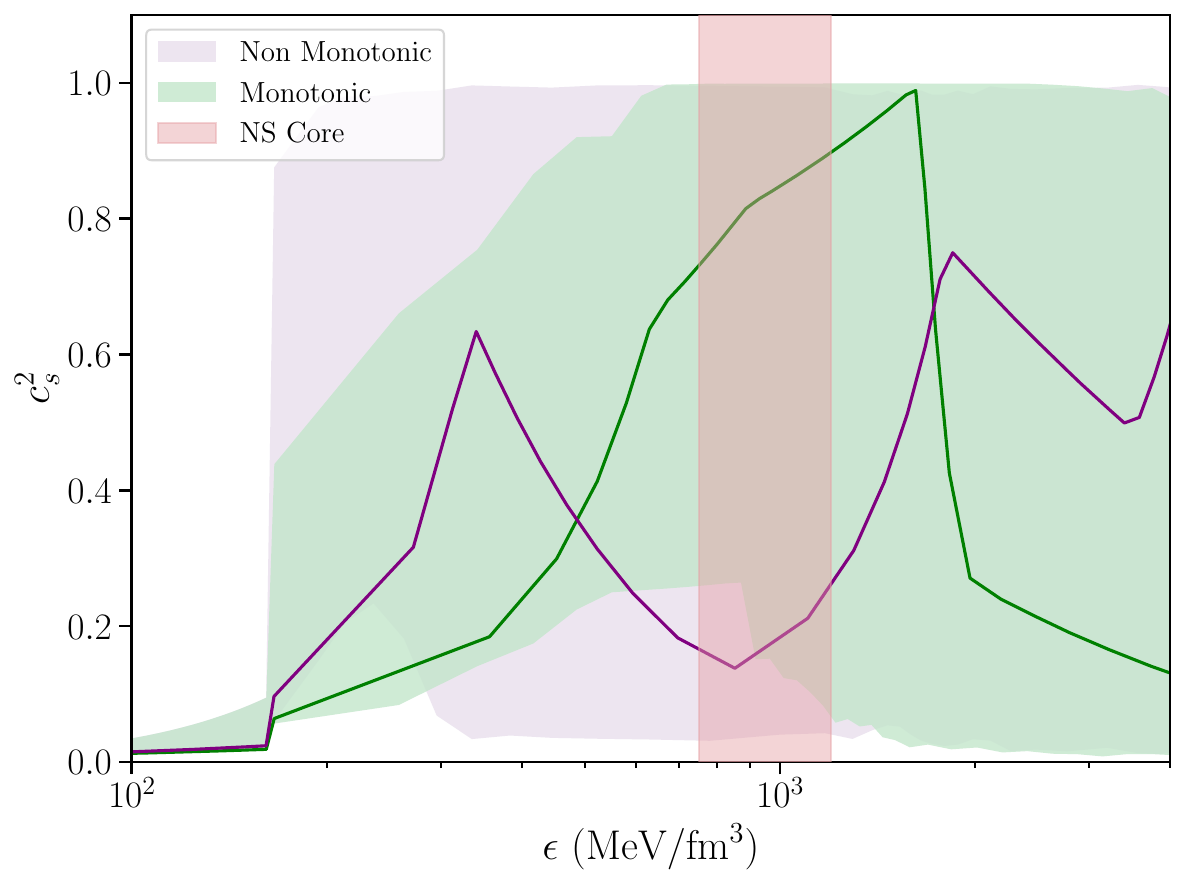}
    \includegraphics[width=0.49\linewidth]
    {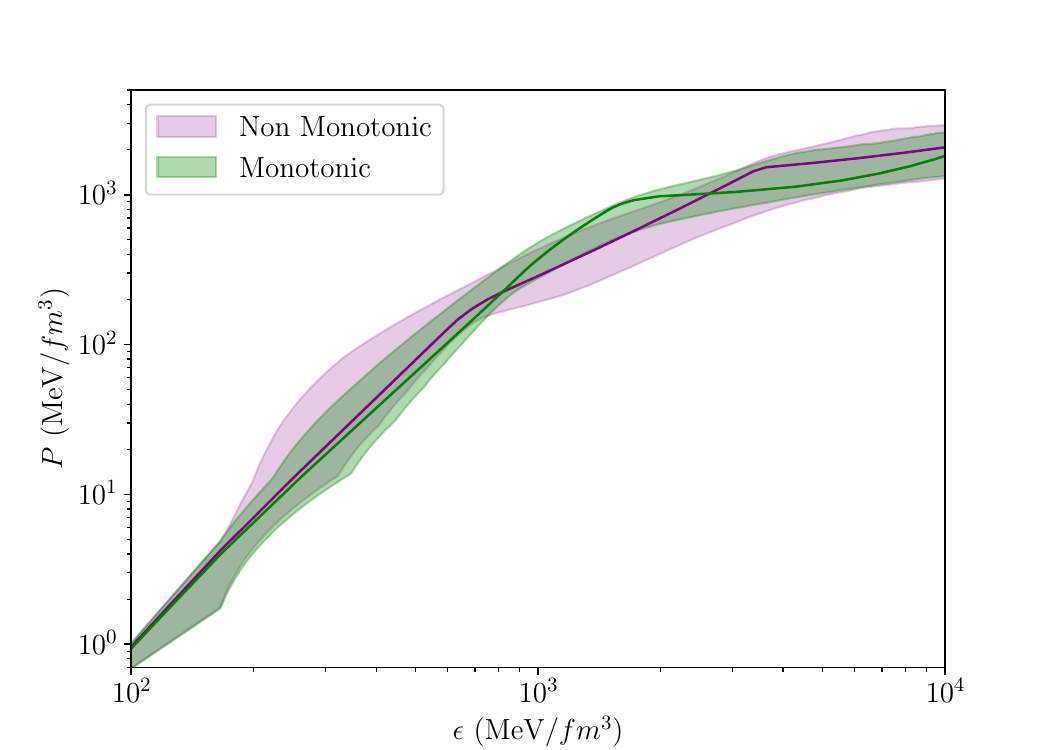}
    \caption{The left panel illustrates the speed of sound square ($c_s^2$)regions spanned by the monotonic (green) and non-monotonic (purple) classes of EOSs. The solid lines represent the specific EOSs, which are used to plot metric potential variation with respect to radial distance in the next figure. The maximum central densities achieved by the maximum mass stars of these sets are also marked in orange.
    The right panel of the plot shows the regions spanned by the monotonic (green) and non-monotonic (purple) EOS classes.}
    \label{EOS}
\end{figure*}

\begin{figure*}
    \centering
    \includegraphics[width=0.48\linewidth]{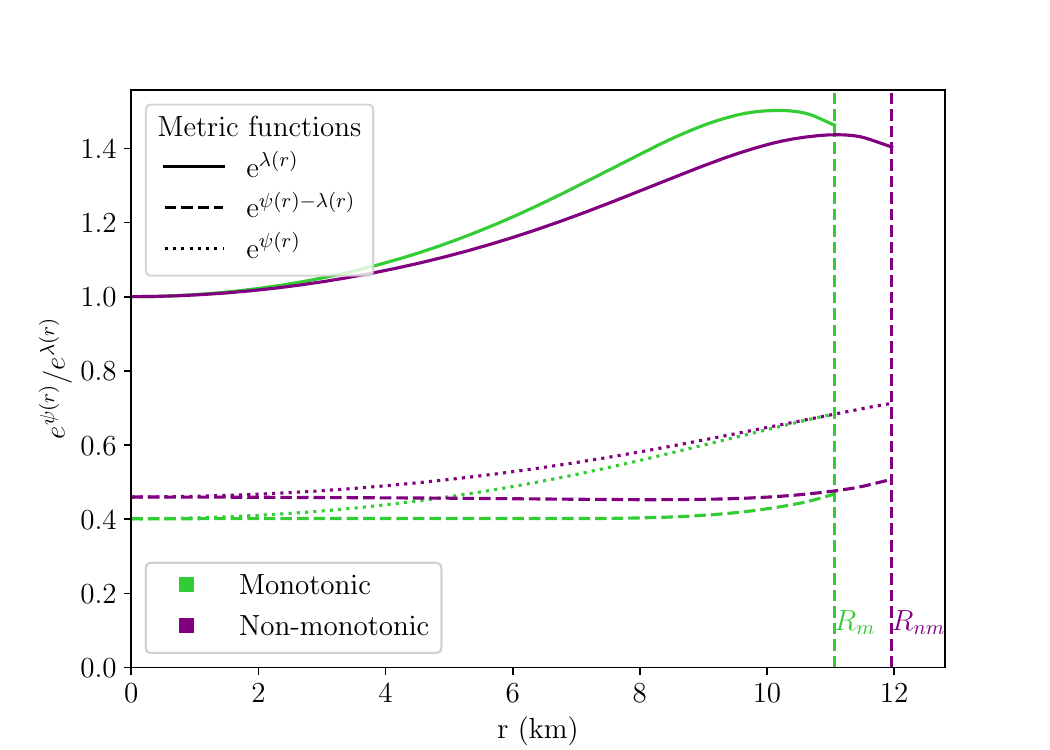}
    \caption{Variation of the metric potentials as functions of the radial distance (r) from the centre of a neutron star, corresponding to the solid-line EoS in Fig. \ref{EOS}. The solid, dotted, and dashed curves represent the metric potentials indicated in the legend. The green and purple curves correspond to the monotonic and non-monotonic EoSs, respectively. All profiles are for 2 $M_\odot$ stars with central energy densities of 980 $MeV/fm^3$ (monotonic) and 820 $MeV/fm^3$ (non-monotonic). The vertical dashed lines denote the radius of the monotonic star (\(R_m\)) and the non-monotonic star (\(R_{nm}\)).}
    \label{pot_cs2}
\end{figure*}

%The solid lines represent the specific EOSs for which the corresponding metric potentials are shown in the right panel.

The speed of sound is an important parameter to characterise the EoS. The change in the stiffness of the EoS is expressed in terms of the sound speed. At sufficiently high density, the EoS of matter is not well understood. Theoretical and experimental limitations arise from the complex problem of understanding QCD at low temperature and high density. At low densities (up to nuclear saturation), the EoS is well constrained by ab-initio calculations \cite{Hebeler:2013nza, Tang:2024jvs}, while perturbative QCD predicts that $c_s^2 \to 1/3$ at asymptotically high densities \cite{Fraga:2013qra, Komoltsev:2022bif}. However, the intermediate density regime, spanning several orders of magnitude between these limits, remains highly uncertain. Theoretical models must ensure that positivity and causality ($0 \le c_s^2 \le 1$) are maintained and that they connect smoothly between the low- and high-density limits. Recent works have used these physical and observational bounds to construct model-independent bands for the sound speed \cite{Annala:2019puf, Annala:2021gom, Annala:2023cwx, sagnik_sos}, which significantly restrict the viable EoS space consistent with all available astrophysical data \cite{Verma_2025, Verma:2025dez}.

\begin{figure*}
    \centering
    \includegraphics[width=0.45\linewidth]{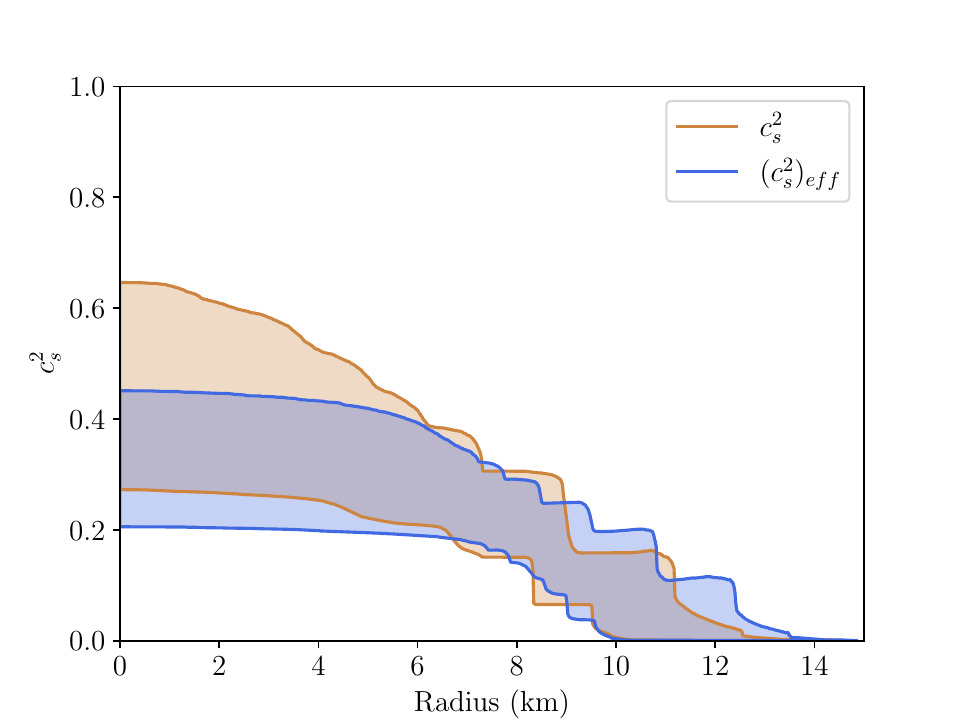}
    \includegraphics[width=0.45\linewidth]{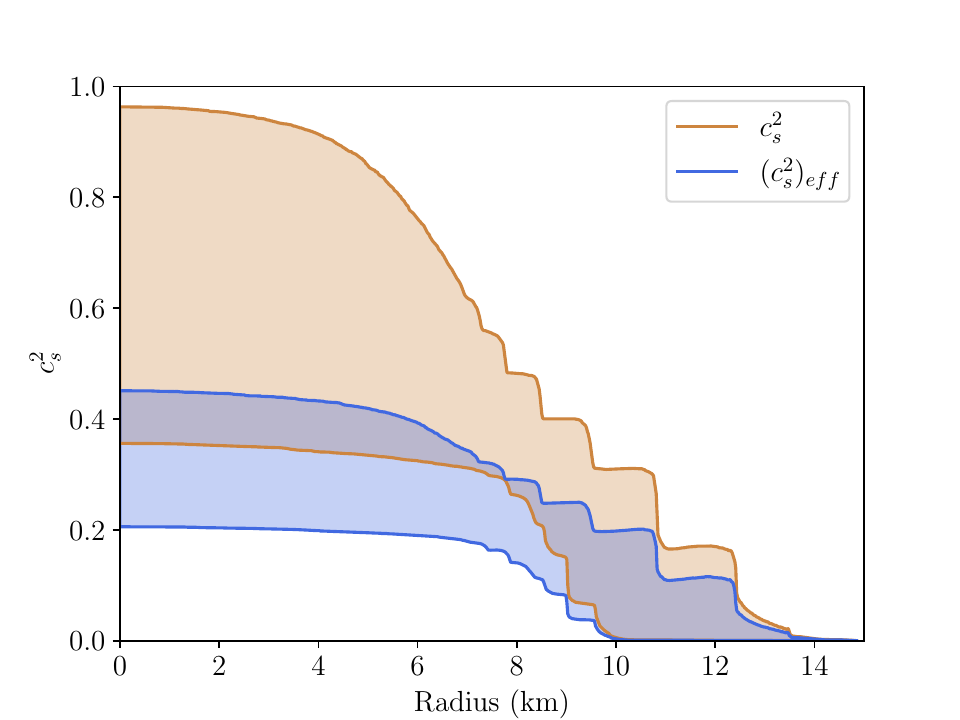}
    \caption{Comparison of \(c_s^2\) and \((c_s^2)_{eff}\) for the monotonic set of EOS. The comparison for the ensemble of 1.4 \(M_{\odot}\) stars is shown in the left panel, while for the 2.0 \(M_{\odot}\) stars is shown in the right panel.}
    \label{cs2_had_comparison}
\end{figure*}

\begin{figure*}
    \centering
    \includegraphics[width=0.45\linewidth]{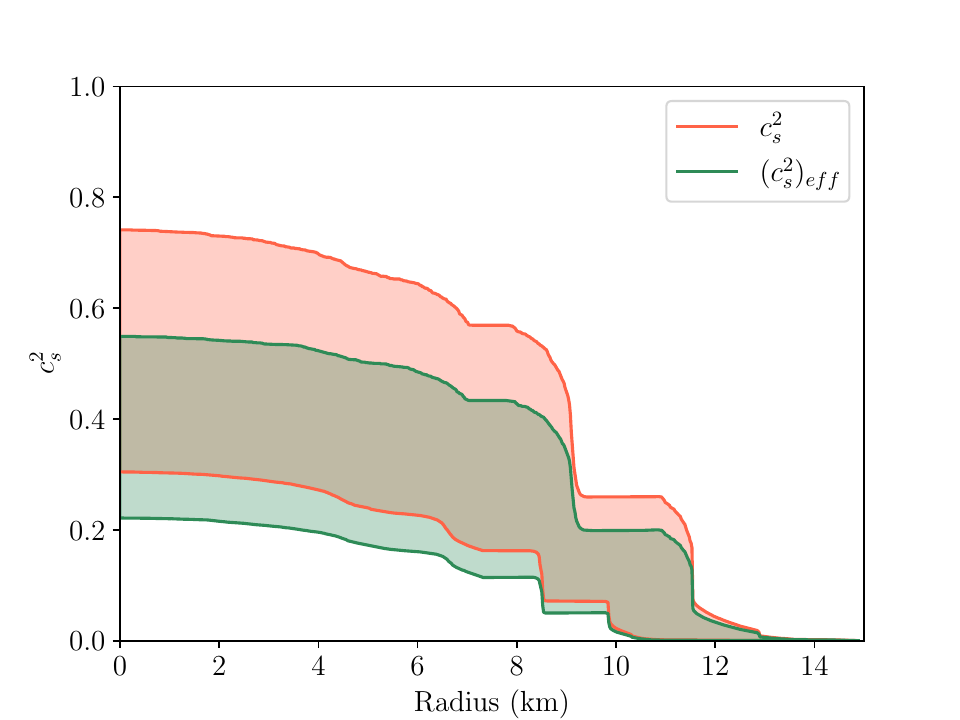}
    \includegraphics[width=0.45\linewidth]{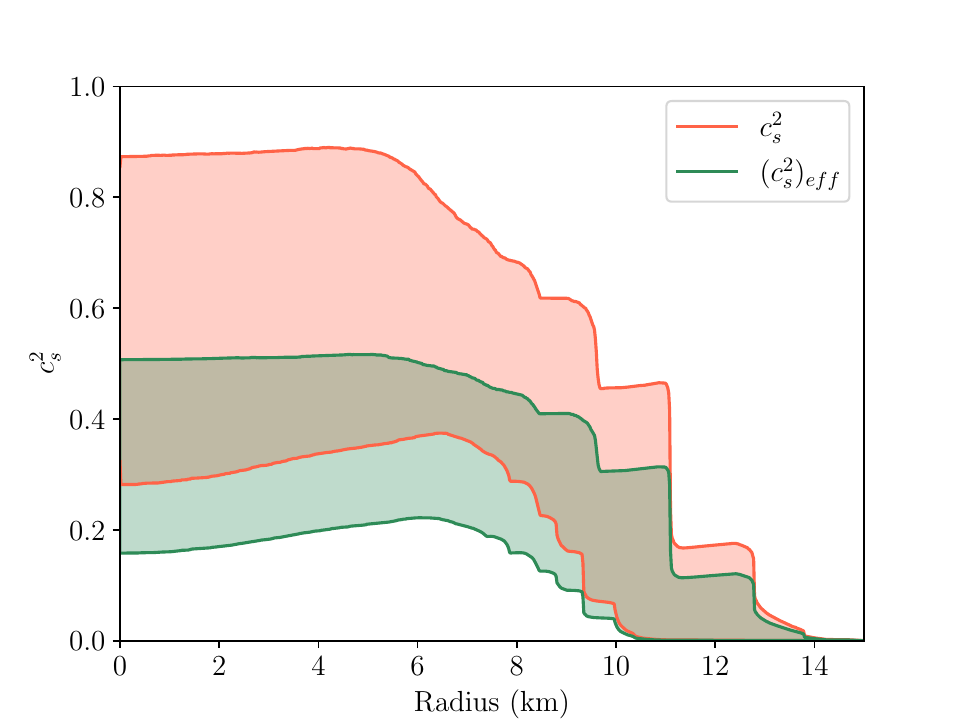}
    \caption{Same comparison of \(c_s^2\) and \((c_s^2)_{eff}\) for the non-monotonic set of EOS.}
    \label{cs2_qua_comparison}
\end{figure*}

The definition of the speed of sound is universal; however, as seen from our calculation, the wave equation is modified near a compact object (under strong gravity), where finite curvature effects are present. One can then define an effective sound speed that governs the transfer of information in a dense medium with finite curvature effects. The matter at the core of neutron stars is under extreme gravity, and information propagation at those densities is expected to be governed by the effective sound speed rather than the usual sound speed. 

\begin{figure*}
    \centering
        \includegraphics[width=0.45\linewidth]{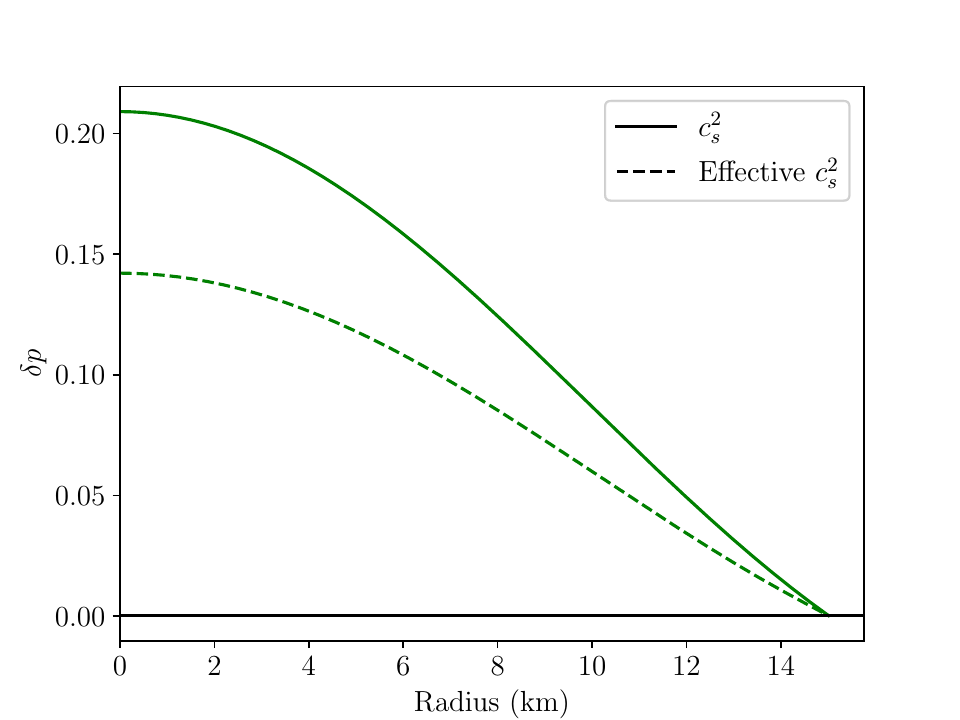}
        \includegraphics[width=.45\linewidth]{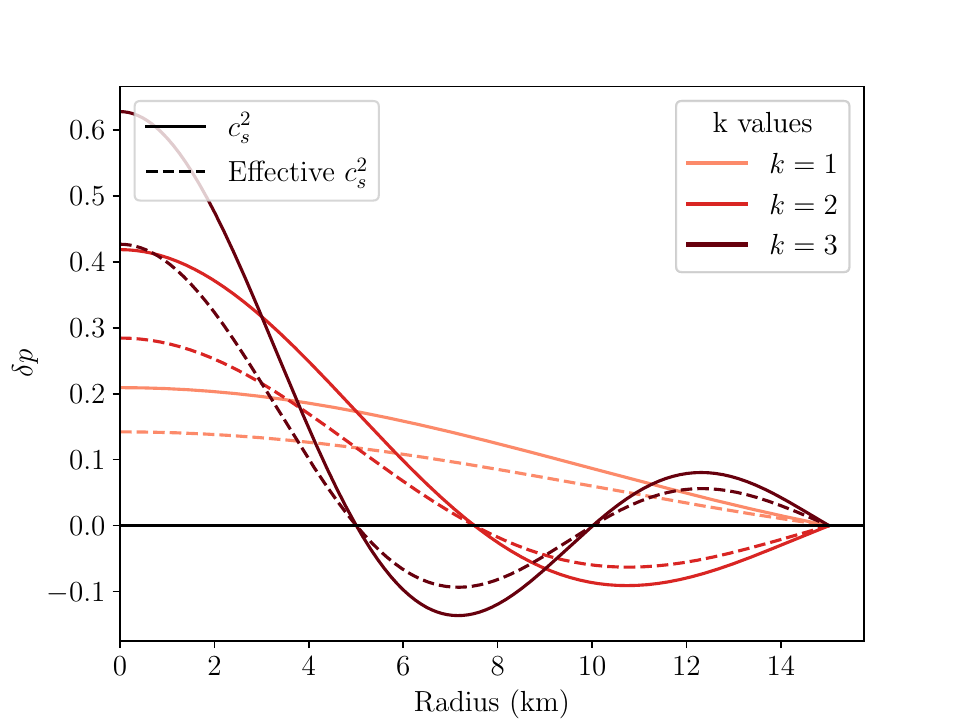}
    \caption{\textit{Left:} Effect of the effective sound speed \((c_s)_{eff}\) on the pressure perturbation (\(\delta p\)) inside the star. The strength of the perturbation is found to be significantly greater than that obtained when considering only the standard sound speed \(c_s\). \textit{Right:} The next three higher modes of the perturbation are shown. The \(k\) values denote the number of nodes within the star. It is also observed that higher modes produce larger perturbations at the centre of the star, indicating the greater energy associated with these modes.}
    \label{Eff_vs_cs2}
\end{figure*}

To analyse the effects of the speed of sound and perturbations inside the star, we construct two sets of agnostic EOS (using sound-speed parametrisation \cite{Annala:2019puf, Annala:2021gom, Annala:2023cwx}) for our study: monotonic and non-monotonic.
The EoSs which have a peak in the sound speed at densities less than the densities corresponding to $M_{TOV}$ are classified as non-monotonic EoS, and those which do not have a peak in the sound speed at densities below the densities corresponding to $M_{TOV}$ are classified as monotonic EoS, as depicted in Fig. \ref{EOS} (left panel) \cite{Verma:2025dez}.
Applying the present astrophysical constraint on the agnostically generated EoS, one identifies the allowed EoS that describes matter at extreme densities in the cores of NSs \cite{boyle_2020, Altiparmak:2022bke, sagnik_sos}.

The thermodynamically consistent EoS contours for the two parameter sets in the ($P-\epsilon$) plane are displayed in Fig.\ref{EOS} (right panel). For each EoS, we examine the corresponding behaviour in the ($c_s^2 - \epsilon$) plane, shown in the left panel of Fig.\ref{EOS}, in order to assess the sound speed across both monotonic and non-monotonic EoS. The shaded region denotes the range of central energy densities that yield the maximum neutron-star mass upon solving the Tolman–Oppenheimer–Volkoff (TOV) equations\cite{TOV_1939}.

To investigate wave propagation in the dense and strongly curved interior of a neutron star, we solve the TOV equations for a given EoS to obtain the spacetime metric potentials, which govern the effective sound propagation. Two representative EoS choices, indicated by solid curves within the EoS contour, are selected for illustration. The resulting metric potentials are shown in the Fig.\ref{pot_cs2}. The radial metric potential ($e^\lambda$) equals unity at the stellar centre (r=0), reflecting local flatness, and increases monotonically outward, reaching a maximum near the stellar surface. This behaviour encodes the cumulative gravitational field generated by the enclosed mass. The temporal metric potential ($e^\Psi$) remains positive and less than unity throughout the star and increases smoothly with the radius of a single star, as indicated by the dotted curves. At the stellar surface, both metric functions are matched continuously to the exterior Schwarzschild solution, ensuring the consistency of the interior–exterior spacetime. The dashed curves in Fig.\ref{pot_cs2} show the ratio of the metric potentials, further used in the calculation of the effective speed of sound.

The quantity \(e^{(\psi - \lambda)}\) is responsible for \((c_s^2)_{eff}\), which remains well below 1 throughout the star. Hence, one could conclude that the \((c_s^2)_{eff}\) is always smaller than \(c_s^2\) inside the star, indicating that strong gravitational effects act to reduce the propagation speed of sound. The comparison for the monotonic and non-monotonic sets of EOS is shown in Fig.\ref{cs2_had_comparison} and Fig.\ref{cs2_qua_comparison}, respectively. 
The results specifically presented for $1.4\,M_{\odot}$ and $2.0\,M_{\odot}$ stars. The region outlines the $90\%$ confidence interval of a large ensemble of EoS for a particular class of EoS. The speed of sound is maximum at the centre of the star for a monotonic EoS as it increases with density. As the density is higher at the core of a $2.0\, M_{\odot}$ star, the speed of sound is also larger. It is evident that $(c_s^2)_{\mathrm{eff}}$ decreases, as indicated by the shift of the $(c_s^2)_{\mathrm{eff}}$ band toward smaller values in all our cases. Moreover, this shift is more pronounced for the more massive stars, indicated by the greater shift for the case of 2.0 \(M_\odot\) stars. The enhanced reduction of \((c_s^2)_{\mathrm{eff}}\) in massive stars arises solely from the metric potential ratio \(e^{\psi-\lambda}\), which deviates more strongly from unity in massive compact objects. As a result, general relativistic effects on sound propagation are significantly weaker in low-mass stars and become increasingly important as the stellar mass increases.

A similar behaviour is also found for stars constructed with a non-monotonic EoS (Fig. \ref{cs2_qua_comparison}). However, there is just one contrasting difference. The non-monotonic EoS has a peak in the sound speed (probabilistically at higher densities, as constrained by NICER), which is evident in the $2.0\,M_{\odot}$ plot. The sound speed reaches its maximum not at the centre but somewhere in between. Therefore, the core of such stars is less stiff than their surrounding layers. The feature is preserved even for the effective sound speed. However, the magnitude is considerably smaller. This makes the star's inner core effectively softer, which can significantly affect the propagation of information from the centre to the surface of the star.

One of the most significant effects of curvature at high density can be understood by looking into how the density (or pressure) perturbation propagates from the core to the surface of the star. This is typically studied using astroseismology \cite{campolattaro-1967, mino_1998, hye-1999, andersson_1996}, which involves examining both radial and non-radial oscillations. The gravitational metric potentials attenuate the oscillation mode frequencies and have been extensively studied in the literature \cite{glendenning2012compact, sotani-2020, zheng-2023, kumar_2025, Thakur_2024}. 

Figure \ref{Eff_vs_cs2} shows the radial perturbations in a \(2.0\,M_\odot\) neutron star described by a monotonic EoS. The fundamental mode, characterised by the absence of radial nodes, is shown in the left panel, while the first two higher overtones are presented in the right panel (\(k=1,2,3\)). The inclusion of the effective sound speed, \((c_s^2)_{\mathrm{eff}}\), leads to a systematic softening of the perturbations, manifested as a reduction in their amplitudes. The suppression induced by \((c_s^2)_{\mathrm{eff}}\) is strongest at the stellar centre, where both the sound speed and spacetime curvature attain their maximum values. The difference between perturbations computed with \(c_s^2\) and \((c_s^2)_{\mathrm{eff}}\) decreases monotonically with radius and becomes negligible near the stellar surface, where the metric potential ratio approaches unity and the sound speed itself is small. Consequently, the collective effect of spacetime curvature on sound propagation diminishes from the core toward the surface. For higher-order modes, the perturbations exhibit a pronounced enhancement in their central amplitudes, reflecting the increasing sensitivity of short-wavelength oscillations to the high-density, strongly general relativistic core of the star.

It should also be noted that vanishing of perturbation is merely a boundary condition we impose to get physical solutions for the eqn \ref{delp}. By imposing this condition, one gets
\begin{equation}
\begin{aligned}
    \sin\!\left(\frac{2\pi r}{\Lambda}\right) &= 0 \\
    \implies \frac{2\pi r}{\Lambda} &= n\pi \\
    \implies \Lambda &= \frac{2r}{n}
\end{aligned}
\end{equation}
where \(n\in Z\), which allows for only specific values of \(\Lambda\) analogous to the quantisation of a particle in a square well potential. Qualitatively, this behaviour can be interpreted to indicate that, since $\Lambda \propto 1/n$, an increase in the mode number $n$ results in a decrease in $\Lambda$. Consequently, the perturbation energy is increased, which is manifested as an increase in the amplitude of the perturbation. A qualitatively similar behaviour is also observed in the non-monotonic case, which is not explicitly discussed in the text.  

\section{Summary and Discussion}\label{Summary_and_Conclusion}

In this article, we demonstrate that the wave equation is modified due to ST curvature near a massive object. We have derived a consistent covariant wave equation that accounts for general relativistic effects. It was also shown that one can define an effective sound speed which involves both medium thermodynamic variables and metric potentials. The ST curvature significantly reduces the sound speed near a compact astrophysical object. 

We have shown how the curvature affects the sound speed and the matter stiffness for NSs. The curvature effectively reduces the sound speed, thereby making the effective EoS at the centre of the star relatively softer. This affects the propagation of information at the core of the star. The pressure oscillation modes remain unchanged, but the effect of the effective sound speed can be easily visualised in the amplitude of the oscillation modes. The curvature significantly reduces the magnitude of the oscillation modes.

The covariant wave equation is the most general and should be effective for analysing sound-speed propagation in compact astrophysical scenarios. One can employ them even to study the propagation of sound or even shock waves, not only near an NS but also near Black holes. It can significantly impact the calculation of jets from massive black holes and shock waves near a supernova.

\begin{acknowledgments}
The authors thank the Indian Institute of Science Education and Research, Bhopal, for providing all the research and infrastructure facilities. AV would like to acknowledge the Prime Minister’s Research Fellowship (PMRF), Ministry of Education, Govt of India, for a graduate fellowship. RM acknowledges the Science and Engineering Research Board (SERB), Govt. of India, for monetary support in the form of a Core Research Grant (CRG/2022/000663). The computations were performed on the BHASKARA and GARGI HPC clusters at IISER Bhopal.
\end{acknowledgments}

\nocite{*}

\bibliography{apssamp}% Produces the bibliography via BibTeX.
%\bibliography{apsrev4-2}% Produces the bibliography via BibTeX.

\vspace{0.1 in}

\section*{Appendix}

The 4-velocity in curved ST is given by
\begin{equation}
    u^{\mu} = \dfrac{dx^{\mu}}{ d\tau}  = \dfrac{dx^{\mu}dt}{dt d\tau}\nonumber 
    \end{equation}    
where, the norm of $u^\mu$ is given by $g_{\mu\nu}u^\mu u^\nu = -1 $.
\\
Christoffel Symbols corresponding to the space-time are:
\begin{align*}
    \Gamma^{t}_{t r} &= \Gamma^{t}_{r t} = \frac{d\psi}{dr}; \Gamma^{r}_{\phi \phi} = -\frac{r e^{-2 \lambda}}{\sin^2\theta};\Gamma^{r}_{t t} = e^{2 \psi - 2\lambda} \frac{d\psi}{dr}\\
    \Gamma^{r}_{r r} &= \frac{d\lambda}{dr}; \Gamma^{r}_{\theta \theta}=-re^{-2\lambda};\Gamma^{\theta}_{\phi \phi} = -\sin\theta \cos\theta \\
    \Gamma^{\theta}_{r \theta} &= \Gamma^{\theta}_{\theta r} = \frac{1}{r};\Gamma^{\phi}_{r \phi} = \Gamma^{\phi}_{\phi r} = \frac{1}{r};\Gamma^{\phi}_{\theta \phi} = \Gamma^{\phi}_{\phi \theta} = \cot \theta
\end{align*}
The local rest-frame fluid is stationary in the coordinate system and moves only through time, with no spatial components. So, the four-velocity takes the form of,
\begin{align*}
    u^{\mu} = (e^{-\psi},0,0,0)
\end{align*}

Since $u^\mu$ only has a time component ($u^t = e^{-\psi(r)}$), the projection tensor will affect only the $t$-component, leaving the spatial components unchanged. So the relevant components of $k^{\sigma\mu}$ are:

\begin{align*}
    k^{tt} &= 0, \quad k^{rr} = g^{rr} = e^{-2\lambda(r)}\\
    k^{\theta\theta} &= g^{\theta\theta} = \frac{1}{r^2}, \quad k^{\phi\phi} = g^{\phi\phi} = \frac{1}{r^2 \sin^2 \theta}
\end{align*}
\end{document}